\documentstyle[12pt,epsf]{article}

\def \dfrac #1#2 {\displaystyle\frac{#1}{#2}}
\def\be{\begin{eqnarray}}
\def\ee{\end{eqnarray}}
\def\bq{\begin{equation}}
\def\eq{\end{equation}}
\def\ben{\begin{enumerate}}\def\een{\end{enumerate}}

\def\prl {Phys. Rev. Lett. }\def\pr{Phys. Rev. }
\def\np{Nucl. Phys. }\def\pl{Phys. Lett. }

\def\roughly#1{\mathrel{\raise.3ex\hbox{$#1$\kern-.75em%
\lower1ex\hbox{$\sim$}}}}

\begin{document}
\begin{titlepage}
%\hfill {\large Preliminary}
\hfill {FTUV 97-50; IFIC 97-53}

\vspace{ .2cm}

\hfill \today

\vspace{.2cm}
\begin{center}
\ \\
{\Large \bf Towards an unified picture of constituent and current 
quarks$\dagger$}
%\\
%\vspace{.1cm}
%of the transversity distribution$\dagger$}
\ \\
\ \\
\vspace{.7cm}
{Sergio Scopetta and Vicente Vento$^{(a)}$}
\vskip 0.2cm
{\it Departament de Fisica Te\`orica, Universitat de Val\`encia}

{\it 46100 Burjassot (Val\`encia), Spain}
 
            and

{\it (a) Institut de F\'{\i}sica Corpuscular, Consejo Superior de 
Investigaciones Cient\'{\i}ficas}
\vskip 0.3cm            
                 and
\vskip 0.3cm                
{Marco Traini}
\vskip 0.2cm
{\it Dipartimento di Fisica, Universit\`a di Trento, I-38050 Povo (Trento),
Italy} 

{\it and Istituto Nazionale di Fisica Nucleare, Gruppo Collegato di Trento}                   
\end{center}
\vskip 0.5cm
\centerline{\bf Abstract}
\vskip 0.2cm

Using a simple picture of the constituent quark as a
composite system of point-like partons, we construct the 
parton distributions by
a convolution between constituent quark momentum distributions
and constituent quark structure functions.
We evaluate the latter at a low
hadronic scale with updated phenomenological
information, and we build the  
momentum distributions using
well-known 
quark models. The resulting parton distributions and structure functions 
are evolved to the experimental scale and good agreement 
with the available DIS data is achieved.  
When compared with a similar calculation using non-composite constituent
quarks, the accord with experiment of the present calculation becomes 
impressive. We therefore conclude that
DIS data are consistent with a low energy scenario
dominated by composite,  mainly non-relativistic constituents
of the nucleon. 

\vskip 0.8cm
\leftline{Pacs: 12.39-x, 13.60.Hb, 14.65-q, 14.70Dj}
\leftline{Keywords:  hadrons, quarks, gluons, evolution, parton distributions, 
structure functions.} 
\vspace{.8cm}

{\tt
\leftline {scopetta@titan.ific.uv.es}
\leftline{vicente.vento@uv.es}
\leftline{traini@science.unitn.it}}
\vspace{0.4cm}
\noindent{\small$\dagger$Supported in part by DGICYT-PB94-0080, 
DGICYT-PB95-0134 
and TMR programme of the European Commision ERB FMRX-CT96-008}.
\end{titlepage}

\section{Introduction}
\indent\indent The theory of hadronic interactions
{\it Quantum Chromodynamics}, ($QCD$) \cite{qcd}, is a theory of quarks 
(antiquarks) and gluons, as has been shown in the asymptotic regime, 
where the interaction can be treated perturbatively \cite{pqcd}. 
At low energies, the idea that
baryons are made up of three {\it constituent} quarks and mesons of a
({\it constituent}) quark-antiquark pair \cite{quarks}, the naive quark model
scenario, accounts for a large number of experimental facts \cite{qm}. 
The quest for a relation between the two regimes, i.e., between the 
{\it current} quarks of the theory and the {\it constituent} quarks of 
the model has an old history \cite{melosh} and, in recent years, 
this search has been the leitmotiv of a considerable research effort 
\cite{ccq}. The fundamental problem one would like to understand is 
how confinement, i.e., the apparent absence of color charges and 
dynamics in hadron physics, is realized.

Detailed quark models of hadron structure based on the constituent quark
concept have been defined in order to explain low energy properties
\cite{drgg,ik}. To proceed from these models to the asymptotic regime, where
deep inelastic scattering (DIS) takes place, a hadronic scale is associated to
the model calculations. The experimental conditions are reached by projecting 
the leading twist component of the observable and evolving according to 
perturbative $QCD$ \cite{pp,jr}. The procedure describes succesfully the gross
features of the DIS results \cite{tr97}. In order to produce a more
quantitative fit different mechanisms have been proposed: {\it valence} gluons,
sea quarks and antiquarks, relativistic kinematics, etc... 
We will show  
that some of these mechanisms appear naturally if we endow the constituent 
quarks with structure, using a procedure already advanced in ref. \cite{tr97}.

It was long ago, at the time that $QCD$ was being proposed, that a procedure, 
hereafter called ACMP \cite{acmp}, was proposed to understand the relation 
between the {\it constituent} quarks and the {\it partons} \cite{feyn}. 
In this approach, which we here explore, {\it constituent} quarks are complex 
objects, made up of point-like partons ({\it current} quarks (antiquarks) and
gluons), interacting by a residual interaction described by a quark model. 
The hadron structure functions are obtained as a convolution of the constituent 
quark wave function with the constituent quark structure function. 
This procedure has been recently revived to estimate the structure function of 
the pion with success \cite{apr}.

In the ACMP approach, each constituent quark is dressed by a neutral cloud of
quark-antiquark pairs and gluons, thus, this scenario supports a confinement
mechanism a la De R\'ujula, Georgi and Glashow \cite{drgg}. A few years earlier
a second approach had been developed \cite{kw}, in which the proton is assumed
to be made out of three valence quarks plus a neutral core of quark-antiquark
pairs and gluons, very much in the spirit of recent developments along the
Manohar-Georgi model \cite{mg}. This duality of approaches has to 
do, in modern language, with the implementation of {\it Chiral Symmetry Breaking} 
(CSB). The naive models \cite{drgg} do not contain spontaneous CSB, 
and this phenomenon
has to be implemented if they are to represent $QCD$ at low energies. But does
it have to be done at the level of elementarity that Kuti and
Weisskopf\cite{kw} proposed and the Manohar-Georgi \cite{mg}
philosophy implies?

Summarizing: the ACMP scheme leads to parton distributions given by a
convolution between constituent quark momentum distributions and constituent 
quark structure functions; we will evaluate the latter at the low energy 
hadronic scale with updated phenomenological information and the former using
well-known non-relativistic quark models of hadron structure \cite{ik,iac};
we will evolve the resulting parton distributions and structure functions to
the experimental scale, to check if  good agreement with the available DIS data
is found. 
\section{The theoretical framework}

In the picture we next explore \cite{acmp}, constituent quarks are themselves
complex objects whose structure functions are described by a set of functions
$\phi_{ab}(x)$ that specify the number of point-like partons of type $b$, which
are present in the constituent of type $a$ with fraction $x$ of its total
momentum. We will hereafter call these functions generically the structure
functions of the constituent quarks.

The functions describing the nucleon parton distributions are expressed in
terms of the independent $\phi_{ab}(x)$ and of the constituent 
probability distributions $u_0$ and $d_0$ as,
\bq
f(x,Q^2)=\int_x^1 {dz\over z} [ u_0(z,Q^2) \phi_{uf}({x \over z},Q^2)
+ d_0(z,Q^2) \phi_{df}({x \over z},Q^2) ]~,
\eq
where $f$ labels the various partons,i.e., valence quarks  
($ u_v,d_v$), sea quarks ($u_s,d_s, s$), sea antiquarks ($\bar u,\bar d, \bar
s$) and gluons $g$.

The different types and functional forms of the structure functions of the 
constituent quarks are derived from three very natural 
assumptions \cite{acmp}:
\begin{itemize} 
\item[  i)]The point-like partons are determined by $QCD$, therefore, quarks, 
antiquarks and gluons;
\item[ ii)] Regge behavior for $x\rightarrow 0$ and duality ideas;
\item[iii)] invariance under charge conjugation and isospin.
\end{itemize}

These considerations define in the case of the valence quarks the following
structure function,

\bq
\phi_{qq_v}({x \over z},Q^2)
= { \Gamma(A + {1 \over 2}) \over 
\Gamma({1 \over 2}) \Gamma(A) }
{ (1-x)^{A-1} \over \sqrt{x} }.
\label{csf1}\eq
For the sea quarks the corresponding structure function becomes,

\bq
\phi_{qq_s}({x \over z},Q^2)
= { C \over x } (1-x)^{D-1},\label{csf2}
\eq
and in the case of the gluons we take

\bq
\phi_{qg}({x \over z},Q^2)
= { G \over x } (1-x)^{B-1}~.\label{csf3}
\eq

The other ingredients of the formalism, i.e., 
the probability distributions for 
each constituent quark, are defined according to 
the procedure of Traini
et al. \cite{tr97}, that is, a constituent quark, $q_0$, 
has a probability
distribution determined by 
    
\bq
xq_0(z,\mu_0^2) =  \frac{m_q}{M} \int d^3p n_q (\vec p)
\delta\left(z-\frac{p^+}{M}\right),
\eq
where $n_q (\vec p)$ is its momentum distribution 
in the corresponding baryonic state. This very intuitive 
expression requires of
a support correction \cite{tr97}.

Our last assumption relates to the scale at which the constituent quark 
structure is defined. We choose for it the so called hadronic
scale $\mu_0^2$ \cite{tr97, grv92}. This hypothesis fixes $all$ 
the parameters of
the approach (Eqs. (\ref{csf1}) through (\ref{csf3})). 
The constants $A$, $B$, $G$
and the ratio $C/D$ are determined by the amount of momentum carried by the 
different partons. We choose, 53.5 $\%$ by the valence quarks and 
35.7 $\%$ by the gluons,
which corresponds to a hadronic scale of $\mu_0^2=0.34$ GeV$^2$ in agreement 
with the parametrization of \cite{grv92}. $C$ (or $D$) 
is fixed according to 
the value of $F_2$ at $x=0$ \cite{acmp},
and its value is chosen again according to \cite{grv92}.
We stress that all these inputs are forced only by the updated
phenomenology. 
The values of the parameters obtained are: $A=0.435$,
$B=0.378$, $C=0.05$, $D=2.778$ and $G=0.135$.
We, here, note that the unpolarized
structure function $F_2$ is rather insensitive to the change 
of the sea ($C$, $D$) and gluon ($B$, $G$) parameters.

To complete the process \cite{pp,jr} the above input distributions are 
NLO-evolved in the DIS scheme to 10 GeV$^2$, where they are compared 
with the
data.

\section{Results}
\indent\indent
We will discuss the results for the proton in two models
for $u_0$ and $d_0$:

\begin{itemize}
\item[ i)] The non relativistic model of Isgur and Karl \cite{ik};
\item[ii)] The algebraic model of Bijker, Iachello and Leviatan \cite{iac}.
\end{itemize}

The parameters of the models
are kept as determined by their authors, which fitted them to static 
properties of hadrons. 

In Figs. 1 through 5 we show the results of the present calculation. Figs. 1
and 2 contain the $u$ valence quark distribution, while Figs. 3 and 4
the $F_2$ structure function. We compare the results of 
the present calculation with those of the same calculation without considering
the constituent quark structure. We stress that no fitting 
of the parameters to approach the
unpolarized data has been done. The new parameters, introduced in the 
definition of the $\phi$ functions, have been solely determined 
by the definition of the hadronic scale and by the assumption of Regge 
behavior at low $x$, whose validity $at$ $low$ $Q^2$, has been recently 
confirmed \cite{abfr}. The procedure has provided us automatically with the 
momentum sum rule and no $ad$ $hoc$ modifications of the model wave functions 
have been necessary. In Fig. 5 we show the gluon distribution, $xg(x,Q^2)$,
at the experimental scale. The impossibility to determine this quantity 
in the extreme quark model scenario  motivated the introduction of  
primordial gluons at the hadronic scale \cite{tr97,grv92}. 

The improvement of all of the results with 
respect to previous calculations is impressive. One should stress, 
that the momentum distribution of the model in ref.\cite{iac}, 
understood as a constituent quark momentum
distribution, leads to an amazing agreement with the data. 

\section{Conclusions}
\indent\indent

In previous work we found the limitations associated with naive quark model
calculations when applied to the explanation of DIS data \cite{tr97}. In that
reference we analyzed several paths to extend the formalism to incorporate the
underlying partonic structure in a natural way. A very appealing scheme seemed
to us incorporating the assumption that constituent quarks are not elementary
\cite{acmp}. We have here explored this scenario. Partons (the quarks, 
antiquarks and gluons of $QCD$) at the hadronic scale are generated by 
unveiling the structure of the constituent quarks. We have seen that 
incorporating this structure in a very physical way improves notably 
the agreement with the DIS data. 
From the point of view of the calculation, we must stress, that no
parameters of the model have been changed with respect to the original 
fit to the low energy properties. The new parameters arising from the
description of the constituent quark structure functions have been adjusted
to describe the input scenario according to the hadronic scale philosophy. 
In this way the sea and gluon distributions are generated in a consistent
way.

Taking into account the almost inexistent fit of parameters, the results are 
surprisingly good for both models  \cite{ik,iac}. In particular the momentum
distribution of ref. \cite{iac} seems to have been  defined to 
fit the DIS data, which is not the case.

A quantitative comparison with the data, indicates that our calculation fails 
to reproduce the high $x$ tail of the experimental curves. Our previous
experience tells us that a probable cause for this 
failure might be the lack of
high-momentum components in the model wave function,i.e., the failure of the
models to take into acount the relativistic motion of the constituent quarks,
an omission which we hope soon to remedy \cite{sv97}. 

The same analysis can be easily performed for the polarized case
and it is in progress\cite{sv97}. The idea being if this 
transformation to the partonic regime from the 
constituent regime is able to account for the so 
called {\it spin crisis} \cite{jm,fr,ar}. 
Moreover, since the method seems to be very predictive, we are
confident that it could be useful to estimate unmeasured
quantities, such as the transversity parton distribution
$h_1$ \cite{rs79}, whose quark model analysis
has been already addressed \cite{modh1}.

The introduction of composite quarks is the most natural way 
to solve the old Melosh problem and to understand some of the  
more recent ones 
\cite{jm,fr}. As many times in physics, the relation between 
constituent and current quarks, has not come from a 
symmetry transformation but from unveiling some underlying structure. 
Constituent and current
quarks are not at the same level of elementarity.

The problem of CSB has not been directly addressed in this paper. CSB 
must be implemented in any low energy model. If the implementation has to 
take place at the level  of the composite constituent quarks or at the level 
of the current quarks is a subject of debate \cite{kmww,forte}. 
Our results seem to imply that it is at the level of the constituent 
quarks that CSB has to be introduced, i.e., we retake the beautiful 
discussion in \cite{acmp}, and confer to Regge behavior a fundamental r\^ole, 
which has to be mantained by the chosen CSB mechanism.

We feel safe to conclude that, the current quarks seen at the parton level 
seem to be embedded in the composite constituent quarks seen at lower $Q^2$. 
An unified picture of current quarks, succesfully describing DIS,  
and constituent quarks, succesfully describing static properties, is 
possible. Work is in progress in that direction.

\section*{Acknowledgements}
\indent\indent
We are grateful to Franco Iachello for a clarifying correspondence regarding 
the model of Bijker, Iachello and Leviatan. This work was finished while one 
of the authors (VV) was visiting the Dipartimento di Fisica dell'Universit\`a 
di Trento. He acknowledges partial financial support from INFN and the warm 
hospitality of the members of the Dipartimento.

\newpage
\centerline{\bf \Large Captions}
\hfill\break
\hfill\break
\hfill\break
{\bf Figure 1}: 
We show the parton distribution $xu_V(x,\mu_0^2)$ obtained 
at the hadronic scale $\mu_0^2 =0.34$ GeV$^2$ 
for a) the Isgur-Karl based model \cite{ik} 
with $36\%$ valence 
gluons at the hadronic scale \cite{tr97} (dot-dashed)
and b) the present convolution approach based on the pure
Isgur-Karl wave functions (long-dashed).
The same distributions, $xu_V(x,Q^2)$, evolved at NLO and at a scale of
$Q^2=10$ GeV$^2$, are given by the dashed and full curves, respectively.  
The fit of the data \cite{lai95} at $Q^2=10$ GeV$^2$ is also shown
for comparison (dots).
\hfill\break
\hfill\break
{\bf Figure 2}:
Caption as in Figure 1 for the model of Bijker, Iachello and Leviatan 
\cite{iac}. Again the corresponding 
valence gluon model is that of ref.\cite{tr97} and the 
data those of ref.\cite{lai95}.

\hfill\break
\hfill\break
{\bf Figure 3}:
We show the structure function $F_2(x,Q^2)$ obtained
by NLO-evolution to $Q^2=10$ GeV$^2$ in the present convolution approach,
using the wave functions of \cite{ik} (full).
The result of \cite{tr97} (dashed) for the same quantity is shown.
The data at $Q^2=10$ GeV$^2$, corresponding to the analysis (without heavy
quark sea) of Lai et al. 
\cite{lai95}, are plotted for comparison (dots). 
 
\hfill\break
\hfill\break
{\bf Figure 4}:
Caption as in Figure 3 for the model of Bijker, Iachello and Leviatan 
\cite{iac}. Again the
corresponding valence gluon model is that of ref.\cite{tr97} and the 
data are those of ref.\cite{lai95}. 

\hfill\break
\hfill\break
{\bf Figure 5}:
We show the gluon distribution $xg(x,Q^2)$ at $Q^2 = 10$ GeV$^2$ obtained with
the present approach for the two models investigated: a) Isgur and Karl
\cite{ik} (dot-dashed) ; b) Bijker, Iachello and Leviatan \cite{iac}
(long-dashed). The data are those of ref.\cite{lai95}.

\newpage
\begin{figure}[h]
\vspace{12cm}
\includegraphics{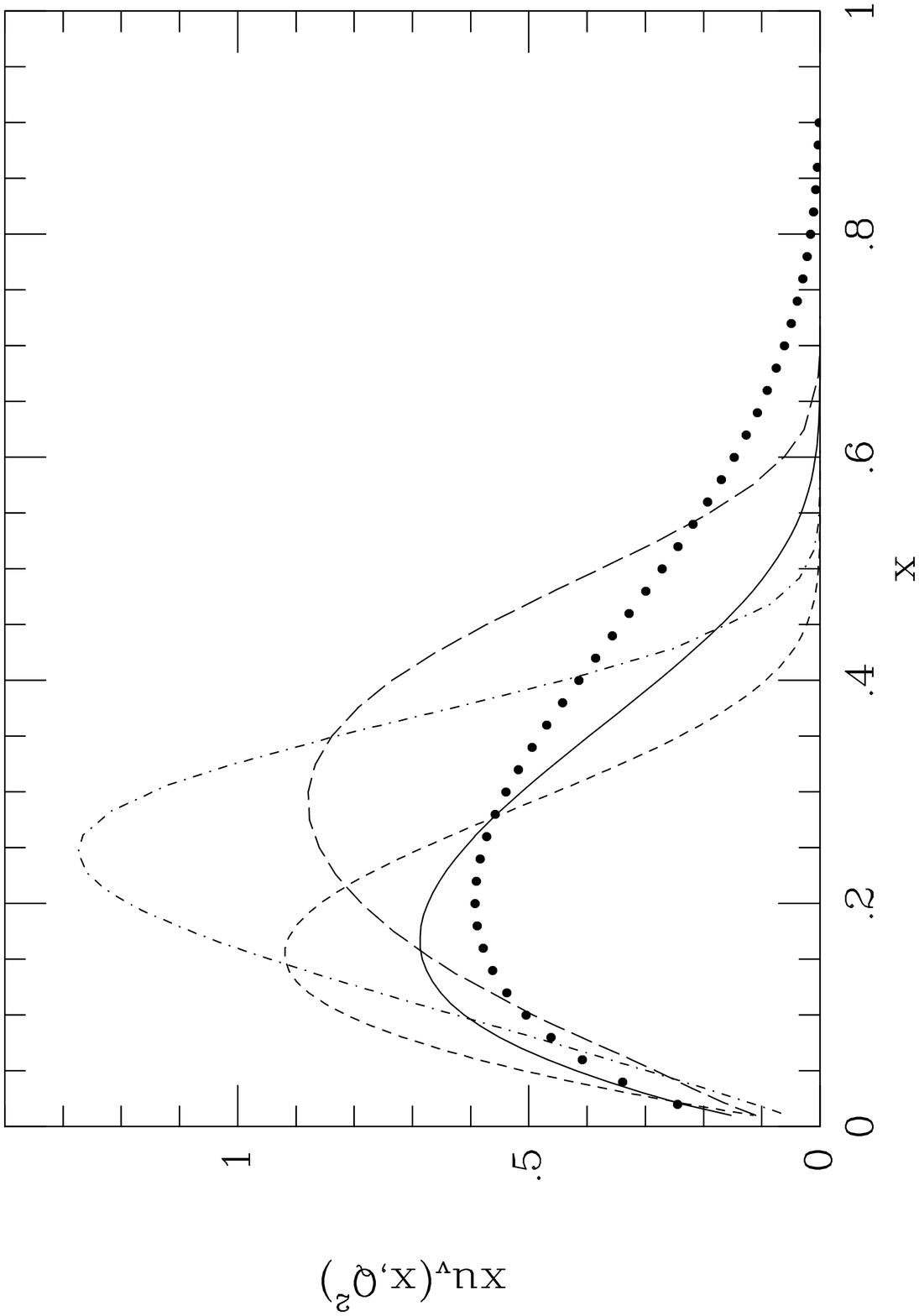}
\end{figure}
\vspace{3cm}
\centerline{\large S. Scopetta, V. Vento and M. Traini}
\vspace{1cm}
\centerline{\bf \large FIGURE 1}

\newpage
\begin{figure}[h]
\vspace{12cm}
\includegraphics{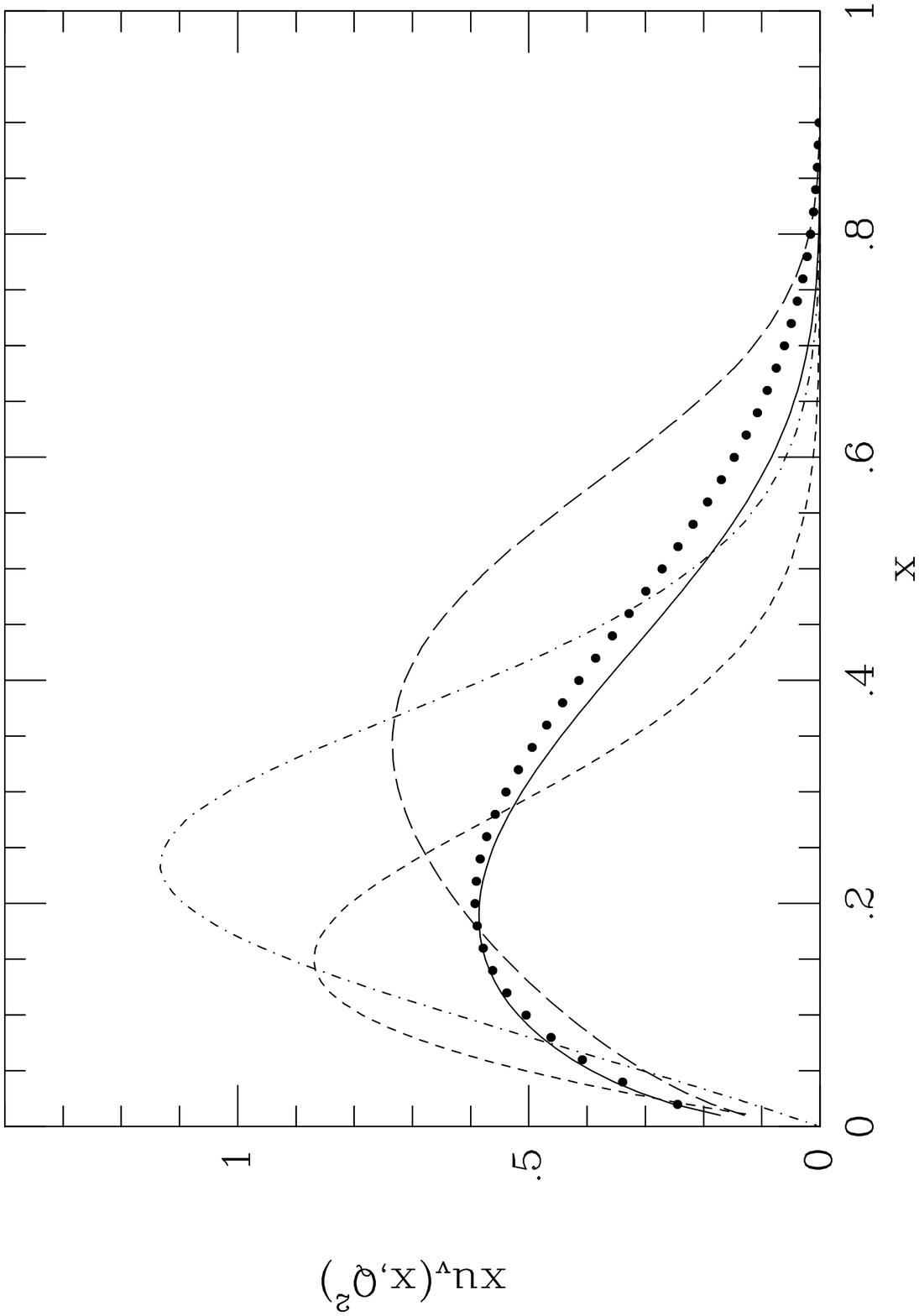}
\end{figure}
\vspace{3cm}
\centerline{\large S. Scopetta, V. Vento and M. Traini}
\vspace{1cm}
\centerline{\bf \large FIGURE 2}

\newpage
\begin{figure}[h]
\vspace{12cm}
\includegraphics{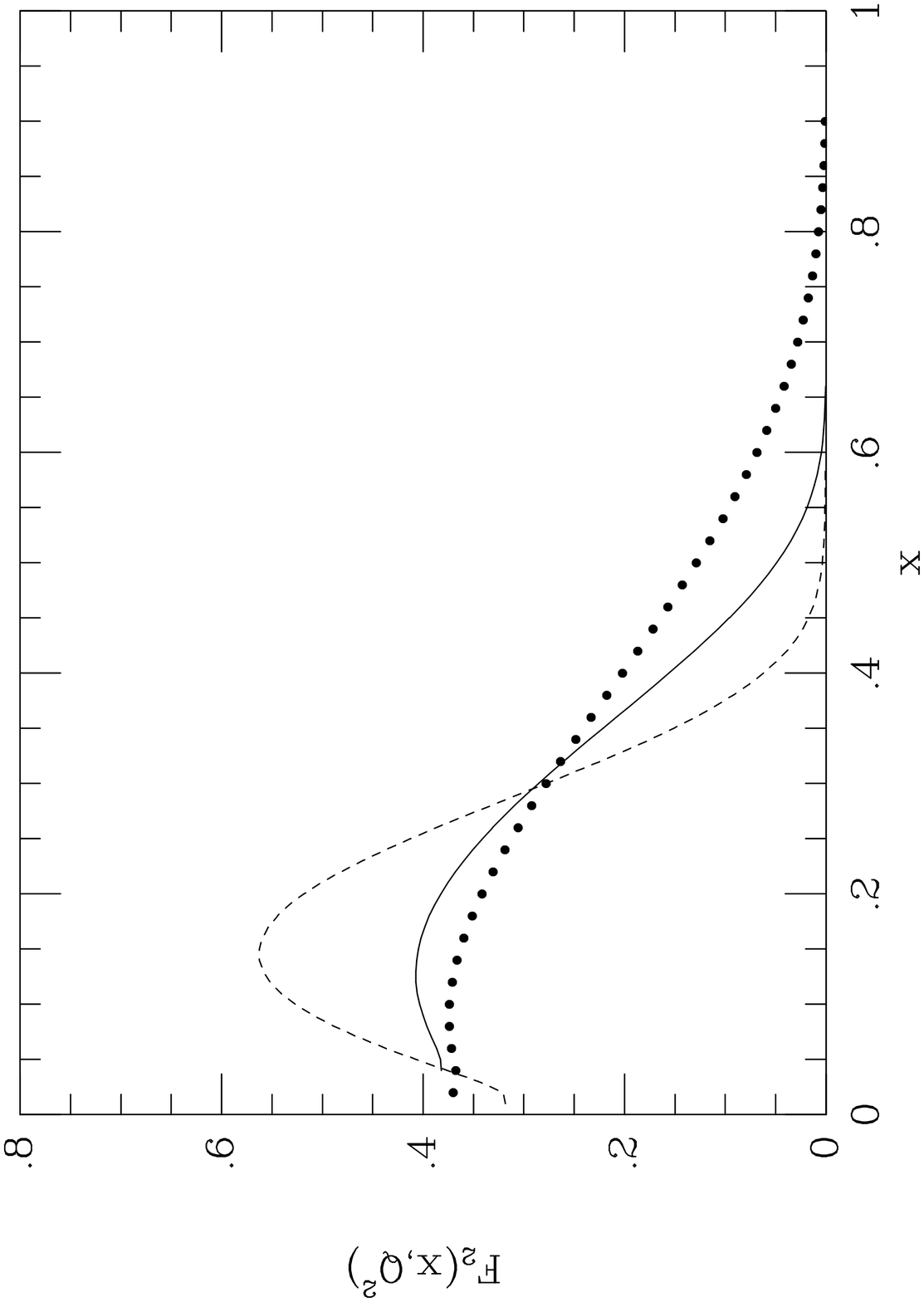}
\end{figure}
\vspace{3cm}
\centerline{\large S. Scopetta, V. Vento and M. Traini}
\vspace{1cm}
\centerline{\bf \large FIGURE 3}

\newpage
\begin{figure}[h]
\vspace{12cm}
\includegraphics{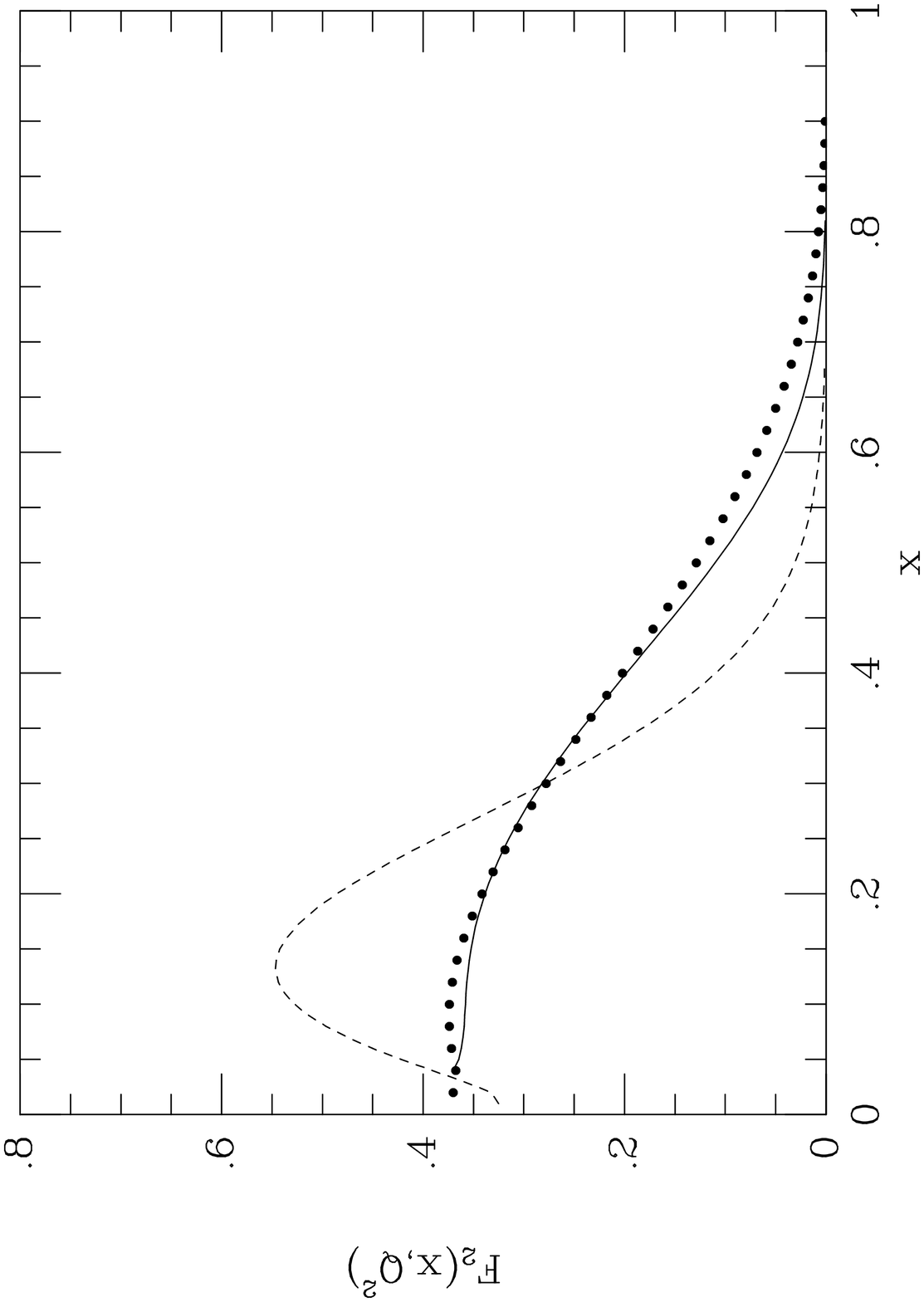}
\end{figure}
\vspace{3cm}
\centerline{\large S. Scopetta, V. Vento and M. Traini}
\vspace{1cm}
\centerline{\bf \large FIGURE 4}

\newpage
\begin{figure}[h]
\vspace{12cm}
\includegraphics{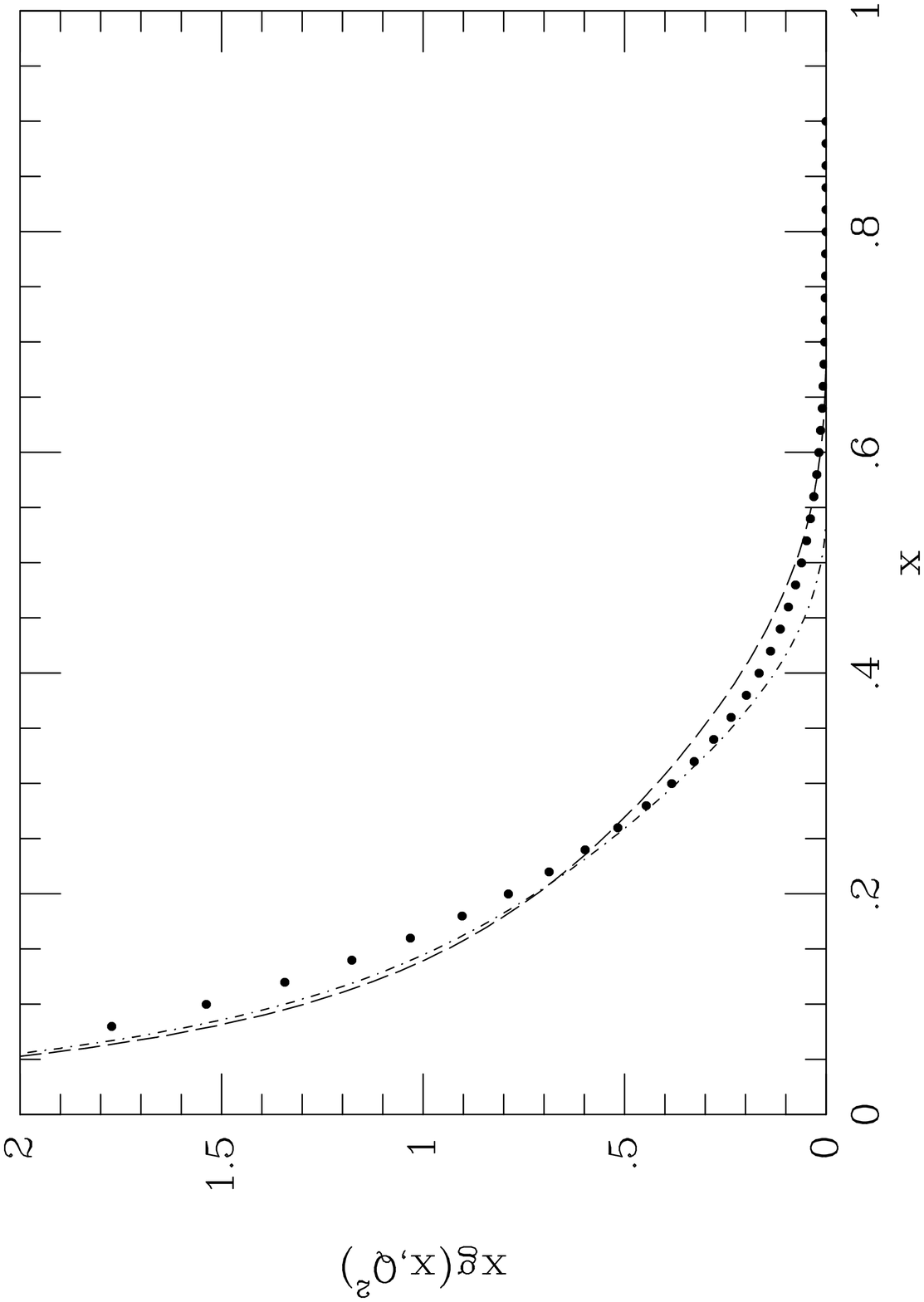}
\end{figure}
\vspace{3cm}
\centerline{\large S. Scopetta, V. Vento and M. Traini}
\vspace{1cm}
\centerline{\bf \large FIGURE 5}

\end{document}